# High-throughput Discovery and Intelligent Design of 2D Functional Materials for Various Applications

*Lei Shen\*, Jun Zhou, Tong Yang, Ming Yang\*, Yuan Ping Feng\**

**CONSPECTUS:** Novel technologies and new materials are in high demand for future energy-efficient electronic devices to overcome the fundamental limitations of miniaturization of current silicon-based devices. Two-dimensional (2D) materials show promising applications in the next generation devices because they can be tailored on the specific property that a technology is based on, and be compatible with other technologies, such as the silicon-based (opto)electronics. Although the number of experimentally discovered 2D materials is growing, the speed is very slow and only a few dozen 2D materials have been synthesized or exfoliated since the discovery of graphene. Recently, a novel computational technique, dubbed "high-throughput computational materials design", becomes a burgeoning area of materials science, which is the combination of the quantum-mechanical theory, materials genome, and database construction with intelligent data mining. This new and powerful tool can greatly accelerate the discovery, design and application of 2D materials by creating database containing a large amount of 2D materials with calculated fundamental properties, and then intelligently mining (via high-throughput automation or machine learning) the database in the search of 2D materials with the desired properties for particular applications, such as energy conversion, electronics, spintronics, and optoelectronics.

In this Account, we summarize our recent progress in the emerging area of 2D materials discovery, database construction, design of 2D functional materials, and device development by materials genome, quantum-mechanical modeling, high-throughput calculations, and machine learning. We developed an open 2D materials database - 2D Materials Encyclopedia (2DMatPedia for short), which includes a variety of structural, thermodynamic, mechanical, electronic, and magnetic properties of more than 6,000 2D materials. Using high-throughput screening and machine learning techniques, we identified exotic 2D materials with desired properties for several applications, such as electrocatalysis, magnetic tunnel junctions, piezo-/ferroelectricity, and solar cells. Our open 2D materials database with high-throughput calculations and proper advanced models will greatly reduce the experimental effort in trial and error, narrow down the scope for both experimental and theoretical explorations, and thus boost the fast and sustainable development in the area of 2D materials.

# 1. INTRODUCTION

Two-dimensional (2D) materials and van der Waals (vdW) heterostructures,[1] a new degree of freedom in the materials space, have gained tremendous interest in the applications of electronics,[2] spintronics,[3] valleytronics,[4] optoelectronics,[5] twistronics[6] and slidetronics,[7] such as (photo)electrocatalysis,[8,9] piezo(ferro)electricity,[7,10] spin(orbit) tunnel junctions,[11,12] and flexible solar cells.[13,14] This extremely fast development of 2D materials and vdW heterostructures is prompted by their various and tunable properties, such as the recently reported unconventional superconductivity in magic-angle stacked bilayer graphene and spin(valley)-polarization in rhombohedral trilayer graphene after 14 and 17 years since the birth of graphene, respectively.[6,15] The other driven force to the rapid development in the field of 2D materials is the discovery of a wealth of new 2D materials by high-throughput experiments[16] and high-throughput calculations.[17-19] Especially, the latter fuels the 2D materials pool by several orders of magnitude through two channels. The first and natural approach is called "top-down" which is inspired by the experimental exfoliation of graphene from layered graphite. Using high-throughput screening on 5,619 experimentally known layered 3D compounds, Mounet *et al.* identified 1,036 candidates that might be easily exfoliable to be monolayer 2D materials.[18] They further developed a free and open 2D materials database "Materials Cloud" to store such a large portfolio of 2D materials and the vibrational, electronic, magnetic and topological properties of 258 compounds.[18] Using the same top-down approach assisted by high-throughput calculations, Choudhary *et al.*, Gjerding *et al.*, and Zhou *et al.*, respectively, also constructed open 2D materials databases "JARVIS",[20] "Computational 2D materials Database (C2DB)"[19] and "2DMatPedia"[17], which includes 812, 1,500 and 2,940 exfoliable 2D materials. Besides the physical top-down approach, we also propose a chemical bottom-up design, which doubles the amount of 2D materials to ~6,000 (see **Scheme1** and details in **Section 2**).[17]

The combination of the exponential increase of computing power with matured, cost-effective and empirical parameter-free density functional theory (DFT) codes promotes the high-throughput DFT calculations, which have shown the success as a powerful tool in the area of new materials discovery, such as the high-throughput-enabled development of the Materials Project Database[21] and Magnetic Topological Materials Database.[22] This calculation-first approach is especially effective in searching the complex heterogeneous catalysts. It can efficiently guide experiments to quickly find high performance catalysts without having to synthesize them first. For instance, Zhong *et al.* identified high performance Cu-Al-base $CO_2$ electrocatalysts from 244 different copper-containing intermetallics from Materials Project Database using high-throughput DFT calculations, which was verified subsequently by experiment.[23] Hannagan *et al.* reported the high-throughput-calculation-led experimental discovery of a single-atom-alloy catalyst for propane dehydrogenation.[24] It is worth noting that the high-throughput calculations in the field of materials science is defined as "an automatic and overwhelming flow from ideas to results", which enables scientists not only to discover thousands of new materials, but also to calculate a variety of properties of them as a part of a single study. It, thus, can generate a huge number of materials-related datasets which fuel interest in the application of machine learning (ML) to further accelerate the discovery and design of new materials.

With the rapid development of high-throughput computational techniques in materials science, our ability to generate materials "big data" has greatly surpassed our capability to analyze it with conventional approaches, underscoring the emergence of the machine learning. ML and its subset deep learning (DL), encompassing statistical algorithms and modeling tools used for finding patterns in high-dimensional data, have been successfully employed in the materials science discipline in the past decade.[25] For instance, ML and DL models have shown their power in predictions of the crystal stability,[26] nanoparticle synthesis,[27] diffraction pattern recognition,[28] battery technologies,[29] and the design of molecules, lead-free perovskites and heterogeneous catalysts.[25,30,31] The key role of ML in materials science is to extract useful information or pattern in the processing-structure-properties-performance relationships (PSPP), which are difficult to establish based on conventional laws of physics and chemistry. Once the materials PSPP relationship is established by ML techniques, ones can design new materials with targeted performance using the inverse design (PPSP).

In our group, we leverage on our core expertise in computational 2D materials as well as high-throughput screening and machine learning techniques to develop the capability in 2D materials informatics and a 2D materials platform, including i) database construction, ii) data analytics, and iii) applications (**Scheme 1**). This platform offers a unique, easy-to-implement, and powerful solution for discovering new 2D materials, designing desired functionality, and exploring their technological applications.

**Scheme 1.** The infrastructure of 2D materials database construction, data analytics, and technological applications.

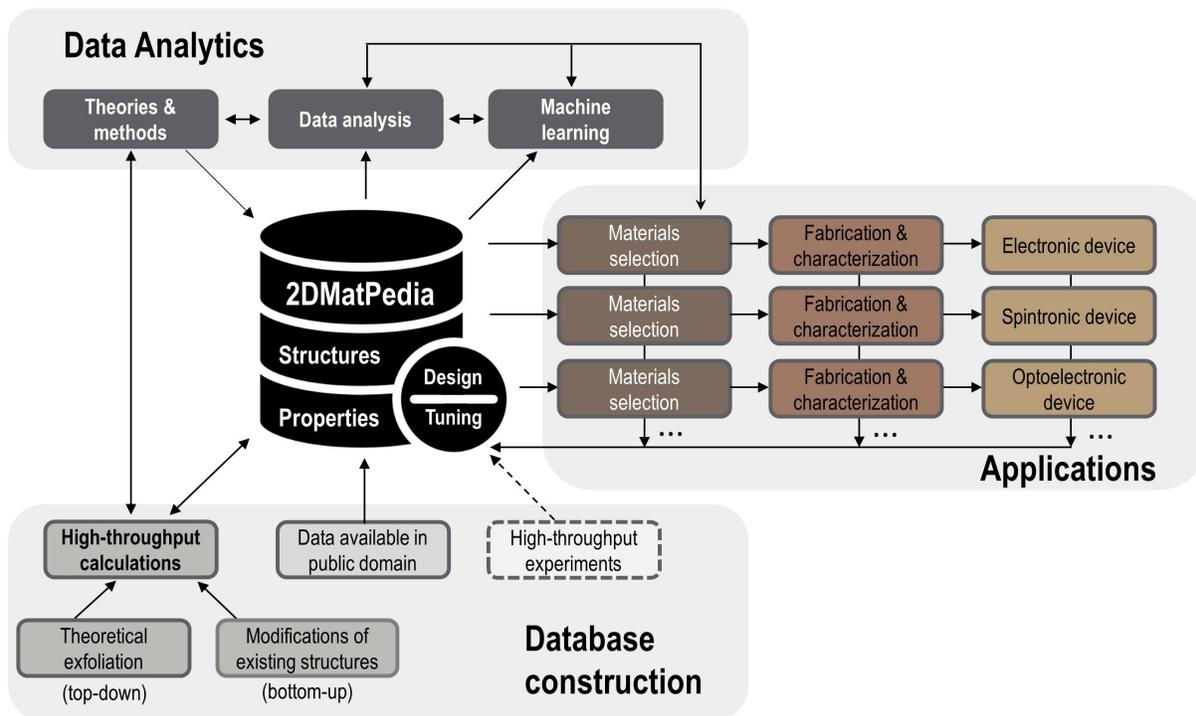

## 2. DATABASE CONSTRUCTION

Using the high-throughput screening of a bulk materials database, we have identified a large dataset of 2D materials, with more than 6,000 monolayer structures, obtained from both top-down and bottom-up discovery procedures.[17] The workflow is summarized in **Fig. 1a**. We started from >80,000 inorganic compounds in Materials Project database (v2018).[21] In the initial stage, we focused on the compounds with simple and layered structures. We then performed high-throughput DFT calculations to calculate the exfoliation energy, optimize these 2D structures, and calculate their properties, which were stored in 2DMatPedia Database. The unary and binary 2D materials obtained from the top-down approach were then used as initial structures for bottom-up elemental substitution. Structure matching was applied to ensure unique structures for further high-throughput DFT calculations.

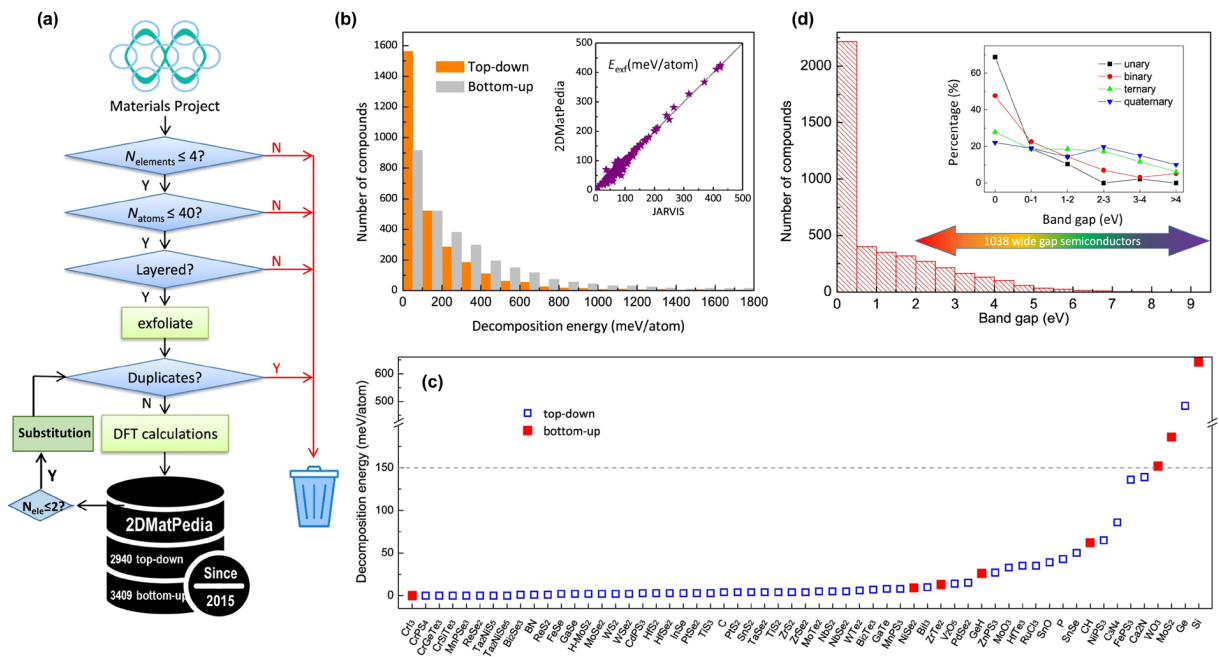

**Figure 1.** (a) The workflow of both top-down and bottom-up design of 2D materials. (b) Statistical analysis of the decomposition energy of 2D materials in 2DMatPedia. Inset shows comparison of the exfoliation energy with that in JARVIS. (c) The calculated decomposition energy for 59 known experimentally grown 2D materials. (d) Statistical analysis of the band gaps for the relatively stable 2D materials in 2DMatPedia. The inset represents the band-gap distribution of different compounds.

In the top-down process, we screened all bulk materials from the database of Materials Project (v2018) for *layered structures* by a topology-based algorithm and theoretically exfoliated them into monolayers. The exfoliation energy, the energy required to isolate a layer from its layered parent, of all layered compounds was calculated and in good agreement with results in JARVIS[20] (**Fig. 1b** and inset). Beside the exfoliation energy, the decomposition energy, the energy needed to decompose a given material into a set of most stable materials (energy above hull), was calculated for investigating the stability of monolayer 2D materials. By checking the decomposition energy of 59 known experimental synthesized monolayer materials, we find that most of them have a low decomposition energy (<150 meV) (**Fig. 1c**). In the bottom-up design process, we generated new 2D materials by chemical substitution of elements in known 2D materials by others from the same group in the periodic table (from H to Bi, excluding radioactive, lanthanide, and actinide elements). We performed high-throughput calculations to obtain the band structures of all 6,000 compounds as shown in **Fig. 1d**. Furthermore, the structural, electronic, magnetic, piezoelectric, electrocatalytic, and energetic properties of these 2D materials were consistently calculated by high-throughput DFT calculations. The details of computational methodology, data record and technical validation can be found in Ref. 17.[17] The whole database is publicly available at http://www.2dmatpedia.org/. This 2D materials-database platform provides a starting point for further material screening, data mining, data analysis and artificial intelligence applications. Some properties and applications of 2D materials discovered based on 2DMatPedia will be discussed in the following sections.

3. INTELLIGENT DATA-MINING 2D FUNCTIONAL MATEIALS

Leveraging the 2DMatPedia database, one can perform intelligent data mining and high-throughput screening to discover functional materials with desired properties for particular applications. To demonstrate this, in this section we first show some examples made in our group, including high-throughput and/or machine-learning screening and design of 2D electrocatalysts, electrides, piezoelectrics, ferroelectric materials, and heterostructure solar cells. Finally, we also summarize recent published works that utilize the datasets in 2DMatPedia database.

3.1 Electrocatalysts

Electrocatalysis is of great importance to the conversion, storage, and utilization of renewable energies for the energy sustainability.[23,25,31] Hydrogen evolution reaction (HER) is the simplest but promising electrochemical process, which is the cathodic half reaction of water splitting and produces hydrogen (an energy-dense clean fuel).[8,32] Nitrogen reduction reaction (NRR) is another important electrochemical reaction for providing ammonia fertilizers from abundant $N_2$ in the atmosphere.[33] The electrocatalysts play a key role in achieving a high conversion efficiency. However, the best electrocatalysts for these two reactions are either noble metal-based or still lacking, especially for NRR at ambient conditions. Among a variety of intensively explored materials, low-dimensional materials (*e.g.*, 0D single atom catalysts[33,34] and 2D materials[8]) stand out, where their profound quantum confinement effect significantly modulates the electronic structures and thus catalytic performance. In addition, the high ratio of exposed surface atoms in such materials greatly improves the atom utilization, which is especially beneficial to the development of noble metal-based electrocatalysts in terms of cost considerations.

**HER of 2D materials**

There is a dilemma in discovering and developing catalysts with 2D materials. Usually, the catalytic performance of experimentally synthesized stable 2D materials is below expectation because of their "stability" (chemically inert). For example, the active sites of $MoS_2$ locate only at the fractional edges, whereas the large basal planes are inert for HER.[35] Thus, the reported methods to improve its catalytic performance are by either maximizing active edge sites or activating the inert basal plane, such as by defect engineering.[36] Both require additional complicated treatments. It is thus desirable to search for 2D materials with intrinsically active basal planes, which could naturally provide a much higher density of active sites. Our 2DMatPedia database includes around 3,000 top-down 2D materials, which provides a wealth of 2D materials for discovering high-performance HER candidates by high-throughput screening[8] or machine learning[37] algorithms.

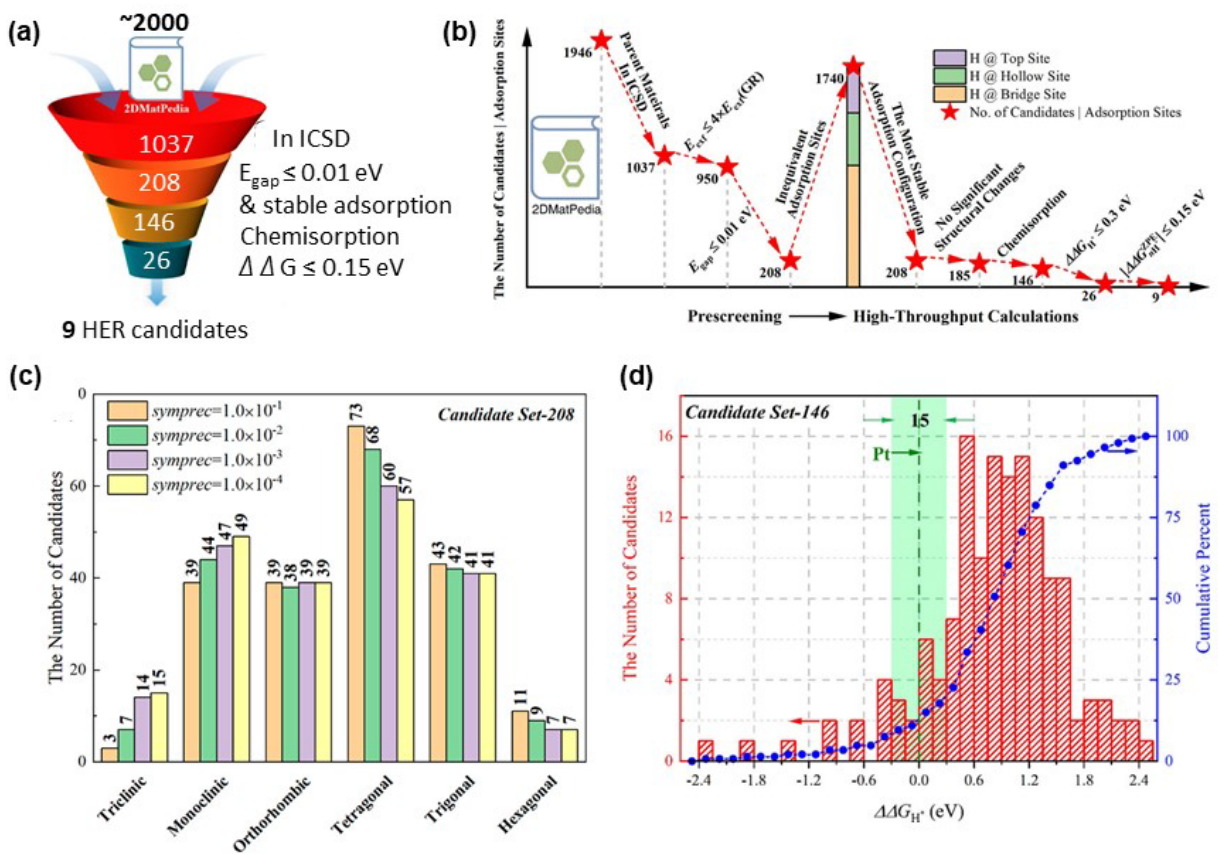

**Figure 2.** (a) Schematic of screening high-performance 2D HER electrocatalysts from 2DMatPedia database. (b) Detailed criteria of prescreening and high-throughput screening. The total number of the considered top-down 2D materials from 2DMatPedia is 1,946. Each candidate set is labeled as Candidate Set-*X*, where *X* is the set size. The stacked column shows the statistics on the inequivalent adsorption sites in Candidate Set-208. (c) Statistical results of Candidate Set-208 at a series of symmetry precisions. (d) Distribution of the hydrogen adsorption Gibbs free energy ($\Delta\Delta G_{H^*}$) with respect to Pt on the catalyst candidates in Candidate Set-146 in the most stable adsorption configuration. The area shaded in light green shows the screened good candidates. Reprinted with permission from Yang *et al*.[8] Copyright 2020 American Chemical Society.

**Figures 2a** and **2b** show the screening processes and the corresponding criteria. We first applied three criteria to prescreen 2D materials in 2DMatPedia: i) their parent layered materials that have been experimentally reported in the Inorganic Crystal Structure Database (ICSD), ii) the upper bound of the exfoliation energy of 268 meV/atom ($4 \times E_{exf}^{graphene}$), and iii) metals with high electrical conductivity. Through the above prescreening criteria, we identified 208 compounds, denoted as Candidate Set-208 for further investigation. Most candidates of this set mainly are of the monoclinic, orthorhombic, tetragonal and trigonal crystal system as shown in **Fig. 2c**. The bridge, hollow, and top adsorption sites of each candidate in Candidate Set-208 are considered, resulting in a total of 1740 adsorption sites. We then performed intelligent high-throughput calculations to automatically iterate over all the possible adsorption sites. Afterwards, these candidates were further screened according to their stability upon hydrogen adsorption as well as the hydrogen adsorption regime (**Figure 2b**), through which 208 candidates were further narrowed down to 146 (Candidate Set-146). **Figure 2d** shows the distribution of the Gibbs free energy ($\Delta\Delta G_{H^*}$) in Candidate Set-146. We finally shortlisted 15 candidates with $|\Delta\Delta G_{H^*}| \leq 0.3$ eV, and 9 best candidates with a more rigorous criterion of $|\Delta\Delta G_{H^*}| \leq 0.15$ eV and the inclusion of the case-by-case zero-point energy contribution. These 9 promising candidates are $IrTe_2$, $Ce_4C_2Br_5$, $C_8$, $Pr_4C_2Cl_5$, $Ba_2Cu_2$, $NbS_2$, $NbSe_2$, $Ti_2Se_2$, and $TaSe_2$ which we call for further experimental investigations. It is worth noting that this 2D HER screening work is keeping updated with the continuous development of 2DMatPedia database.

**NRR on 2D materials support**

Single atom catalysts (SACs) have recently been intensively explored for electrochemical nitrogen reduction reaction because of the 100% utilization, high activity, and high selectivity, in which N-doped graphene is widely used to support the SACs.[34,38] Through high-throughput DFT screening, we found that monolayer $MoS_2$ is a potential alternative substrate to anchor transition metal (TM) SACs towards NRR.[33] It is found that the single Mo atom anchored on top of the Mo site of $MoS_2$ has the best NRR performance (the lowest limiting potential and best selectivity against HER) among Ag, Au, Co, Cr, Cu, Fe, Mn, Mo, Ni, Pd, Pt, Rh, Ru, Sc, Ti, V, W and Zn. The estimated limiting potential of Mo@$MoS_2$-M is around -0.44 V via the distal mechanism. Our results further confirm the stability of the Mo@MoS$_2$-M as well as its good selectivity to NRR against hydrogen evolution reaction.

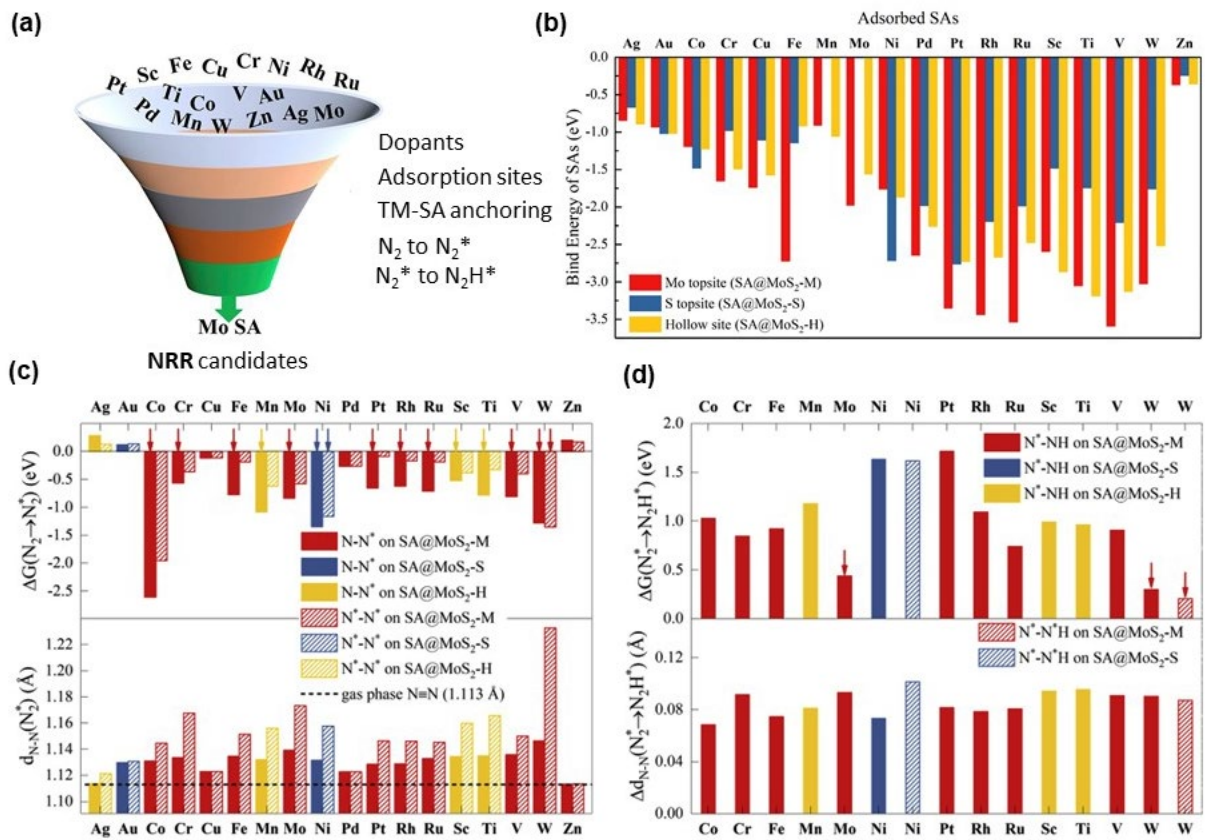

**Figure 3.** (a) Schematic of screening steps of 18 TM-SAs@MoS$_2$ for NRR. (b) The calculated binding energy of 18 TMs on monolayer MoS$_2$ at 3 different adsorption sites. (c) The Gibbs free energy change of N$_2$ adsorption and the N-N bond length of the adsorbed N$_2$ on various TM-SAs@MoS$_2$. The configurations labeled by arrows are selected to proceed the next NRR step - hydrogenation of adsorbed N$_2^*$. (d) The changes in the Gibbs free energy and the N-N bond length upon the hydrogenation of adsorbed N$_2^*$. Mo and W SAs (red arrows) are selected for further evaluation of their NRR activities. Reprinted with permission from Yang *et al*.[33] Copyright 2020 Elsevier.

**Figure 3a** shows the workflow of our high-throughput screening process for NRR. Eighteen transition metals are selected for SACs on three different sites (Mo top, S top and hollow) of monolayer MoS$_2$. Based on the high-throughput binding-energy calculations (**Fig. 3b**), we identified the most stable anchoring site on MoS$_2$ for each transition metal single atom. Since the $N_2 \rightarrow N_2^*$ is the first reaction step and $N_2^* \rightarrow N_2H^*$ is one of the potential limiting steps in NRR, we first calculated the Gibbs free energy change of these two steps as the screening criteria (**Figs. 3c and 3d**). Based on the result, we shortlisted two promising candidates, Mo@MoS$_2$-M and W@MoS$_2$-M (**Fig. 3d**). Through further systematic investigations on the stability, activity, and selectivity against HER, Mo@MoS$_2$-M is finally identified as the best NRR catalytic system.

In a short summary, high-throughput screening approach is a powerful tool to discover 2D materials having good electrocatalytic performance, or enabling synergistic catalysis with the nano-catalysts as the support. Beside monolayer 2D materials, a rational design, such as the heterostructure and Janus structure, can effectively improve the catalytic performance.[9]

3.2 Layered electrides and electrenes

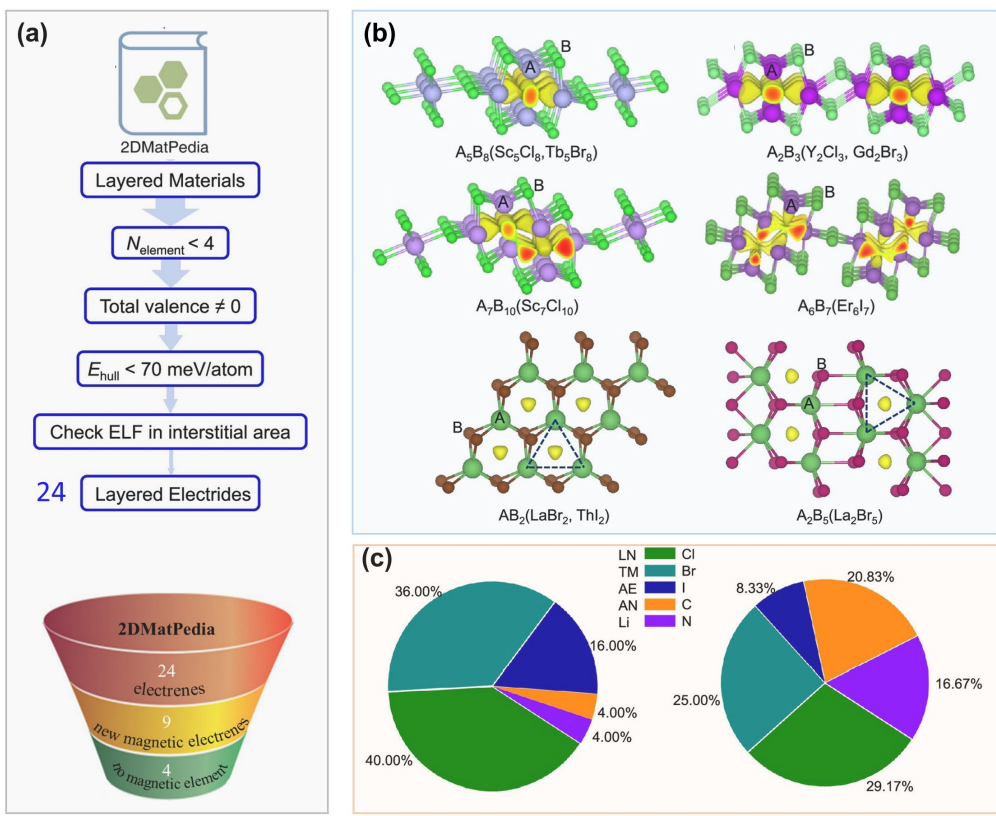

**Figure 4.** (a) Workflow of the high-throughput screening procedure for layered electrides. The decreasing size of light-blue arrow indicates the number of candidates after each screening step. Through the process, 24 layered electrides were identified, and 4 among them are ferromagnetic without magnetic elements. (b) Six types of new structure prototypes of nine new electrides which have never been reported before. (c) Statistical information for the 24 layered electrides screened in this work. Cation (left panel) and anion (right panel) distribution in percentage, respectively. Reprinted with permission from Zhou *et al*.[39] Copyright 2019 American Chemical Society.

Electrides or electrenes in monolayer are intrinsic electron-rich materials, in which excess electrons (named anionic electrons) are confined in geometrically interstitial sites and loosely bounded, acting as anions.[39,40] It has been demonstrated that electrides exhibit many unique properties and have potential applications, such as superconductivity, topological matters, and catalysis. Despite their exotic properties, electrides, especially layered electrides, are of scarcity with only a few compounds grown experimentally. By high-throughput searching layered materials with anionic electrons in 2DMatPedia, we identified 24 layered electrides (**Fig. 4a**).[39] Nine of them have unique compositions which have never been reported before (**Fig. 4b**). Statistic information for the 24 layered electrides shows that the ions are mainly from early transition metal, alkaline earth metal, lanthanide, and actinide (**Fig. 4c**). Interestingly, we found nine magnetic electrenes. To our surprise, four of them ($LaBr_2$, $La_2Br_5$, $Sc_7Cl_{10}$, and $Ba_2LiN$) contain no magnetic elements.[40] The underlying physics of long-range ferromagnetism in electrenes is unrevealed by low-energy effective Hamiltonian models.[40] These screened materials significantly increase the number of layered electrides and 2D ferromagnetic materials, and expand the exploration scope and depth in the area of electrenes and ferromagnetic materials.[41,42]

3.3 Two-dimensional ferromagnets

Magnetism has been explored in 2D materials for more than a decade (see **Ref. 12**[12] and references therein). In 2017, the first observations of intrinsic ferromagnetism in pristine 2D crystals were reported in $Cr_2Ge_2Te_6$ and $CrI_3$.[42] Soon after, a variety of 2D (anti)ferromagnetic materials, such as $Fe_3GeTe_2$, $VSe_2$, $MnSe_2$ and $MnBi_2Te_4$, have been reported experimentally (see review **Ref. 42**[42] and references therein). Most reported 2D magnets only work at low temperatures. However, this does not fundamentally exclude the possibility of high temperature 2D magnets. Efforts using high-throughput DFT calculations and machine learning toward this goal have shown promise.

We have carried out high-throughput spin-polarized calculations to all 2D materials in 2DMatPedia,[17] and identified 230 compounds with magnetic moment above 0.1 $\mu_B$ per unit. By requiring at least one magnetic element (Gd, Fe, Cr, Mn, Co, Ni, V, Rb), and at least one heavy element (Z>49 ensuring large SOC) per formula unit, we then selected 83 candidates for further magnetic coupling calculations within ferromagnetic and several antiferromagnetic configurations. We finally identified 40 ferromagnetic candidates. Their Curie temperature is obtained by an analytical solution based on the 2D Ising model, and 9 of them may be ferromagnetic at room temperature. Utilizing the magnetic information above as the datasets, we developed machine-learning models to classify the 2D materials into the respective categories of non-magnetic and magnetic materials with an accuracy of 93.4% via the gradient boost algorithm. The latter is further classified into anti-ferromagnetic and ferromagnetic materials with the accuracy of 94.4% using the random forest classier. A web-based AI application for predicting 2D magnetic materials has been developed and can be found at https://twodferromagnetism-model.herokuapp.com.

Such 2D ferromagnetic materials informatics, which agrees well with recent experimental reports, is expected to foster practical applications of 2D magnetism, such as 2D spintronic or spin-orbitronic devices.[12] Spin-valley locking and spin-valley polarization has been reported in 2D materials (e.g., $MoS_2$) on magnetic substrates.[43] Magnetic tunnel junctions with 2D magnets as tunneling barriers exhibit giant tunneling magnetoresistance.[11] New concepts of spin field-effect transistors based on 2D magnets have been reported as well. Furthermore, the exotic spin textures, quantum phases, and quasiparticles in 2D magnetic materials and hetero-interfaces can lead to new ways of computation and communication. We envision those successive breakthroughs in 2D magnets could usher in a new era of information technologies with exciting applications in computing, sensing, and memories.[2,3,44,45]

### 3.4 2D Piezoelectrics

Piezoelectric materials enable the interconversion between mechanical and electrical energy. This is made possible by the change in polarization of the material when it is stretched or compressed. As such, piezoelectric materials are integral components of intelligent, multi-functional devices and drive a multi-billion dollar industry through their applications as sensors, actuators, energy harvesters, etc. The recent thrust toward flexible nanoscale devices creates a need for 2D piezoelectric materials, such as wearable sensors and smart material applications that require a large voltage signal in response to a small amount of physical deformation.

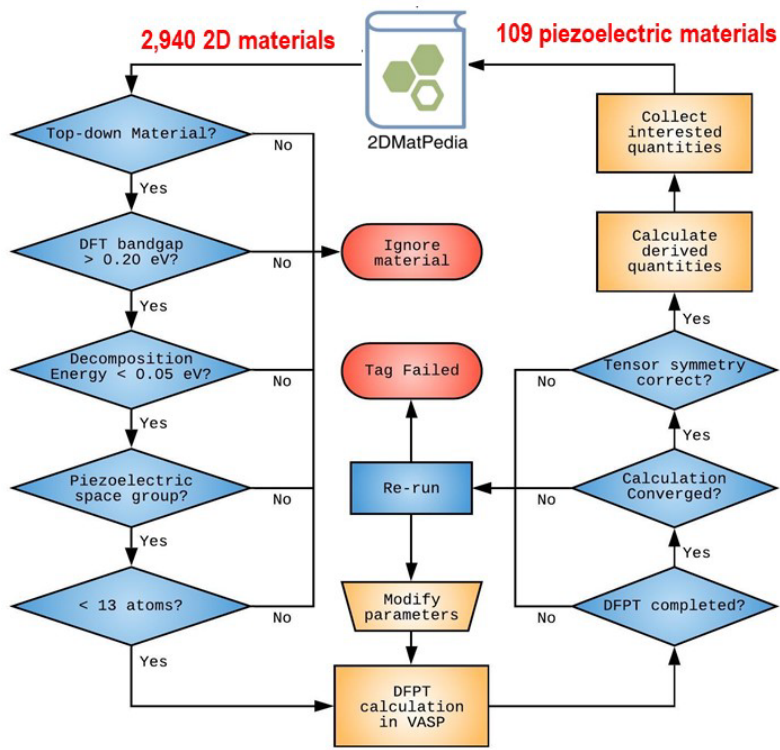

**Figure 5.** (a) Workflow of the high-throughput screening to identify piezoelectric 2D materials. Several criteria were chosen to ensure the stability and existence of dipole. 109 candidates are identified from 2,940 2D materials.

So far, the discovery of 2D piezoelectric materials is mostly by trial and error or materials intuition, for example, by searching on specific 2D ferroelectric materials because ferroelectrics must be piezoelectric. However, there is no direct relationship between the piezoelectric coefficient and ferroelectric coefficient, and a material with large ferroelectric polarization may have very small piezoelectric coefficient. Experimentally, it is also challenging to quantitatively compare the piezoelectric coefficients of 2D materials. To address this problem, we performed a systematic high-throughput search through the 2DMatPedia database to discover 2D piezoelectric materials and identify candidates with large intrinsic piezoelectric coefficients.[10] **Figure 5** shows the workflow with screening criteria. 109 2D piezoelectric materials that we identified from 2,940

"top-down" 2D materials in 2DMatPedia. Interestingly, 46 piezoelectric materials show out-of-plane piezoelectricity. Our findings provide a platform for the development of flexible nanoscale piezoelectric devices, such as high precision actuators and wearable electronics or energy-harvesters.

## 4. RATIONAL DESIGN OF VAN DER WAALS HETEROSTRUCTURES

Combination of two or more different monolayers into one vertical stack via van der Waals forces, i.e., 2D heterostructures, not only tremendously increases the number of 2D materials, but also brings a variety of new properties due to the synergetic or coupling effect, which do not exist in each constituting material.[1] Such heterostructures have already led to numerous exciting properties and a range of applications, such as field-effect tunneling transistors, plasmonic devices, and light-emitting diodes.[1]

To identify van der Waals vertical heterostructures with high power conversion efficiency (PCE) for the target application of flexible solar cells, we conducted a high-throughput screening (**Fig. 6a**) to the band alignments of 1,540 vertical heterostructures formed by 56 2D semiconducting materials (**Fig. 6b**).[13] Based on the band alignment of two different semiconducting compounds (**Fig. 6c**), we identify more than 90 heterostructures with estimated PCE being higher than 15%, of which 17 heterostructures (labelled in **Fig. 6d**) are predicted to have PCE higher than the best value (20%) reported or proposed in the literatures. Taking this example, we demonstrate that van der Waals heterostructures not only increase the members of 2D materials family, but also realize interesting phenomena for a variety of applications. More 2D van der Waals heterostructures and their properties will be provided in 2DMatPedia.

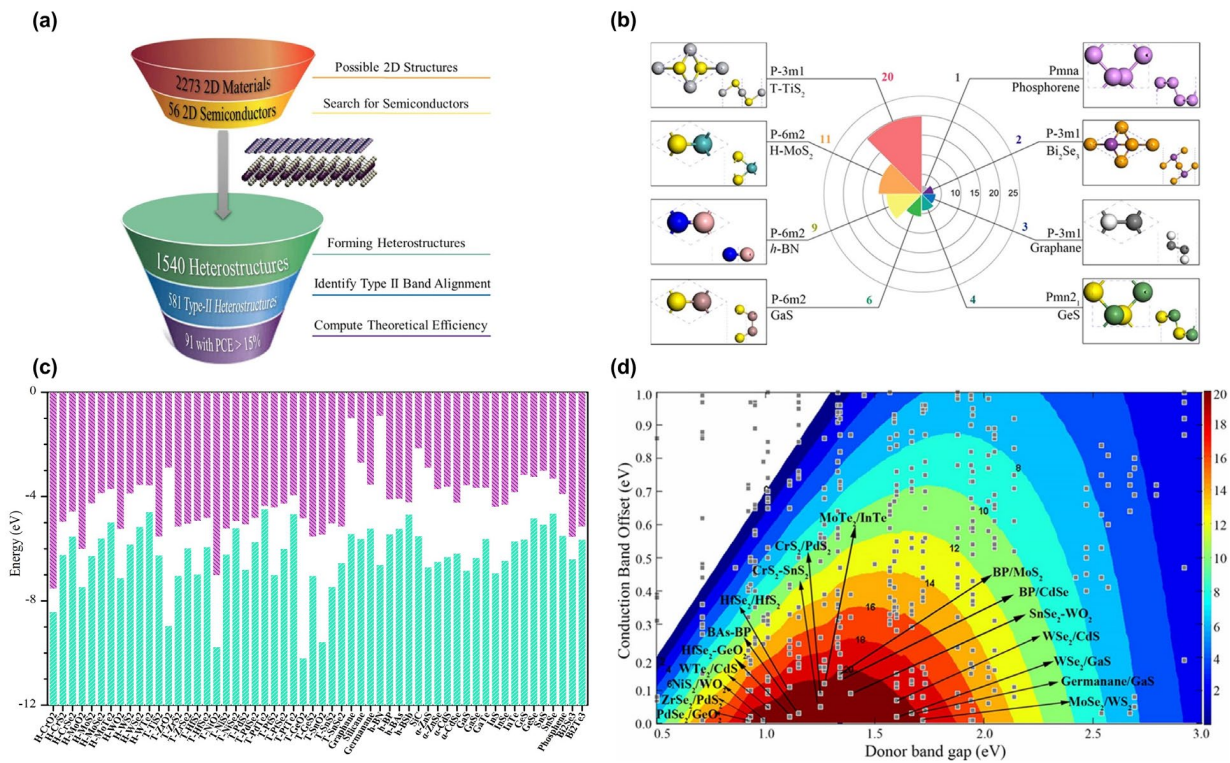

**Figure 6.** (a) Workflow of the design of heterostructures and high-throughput screening criteria for high-efficiency type-II solar cells. (b) Polar histogram of the eight categories with their space group, representative material and the number of structures involved. (c) Band alignment of all of the 2D materials with respect to the vacuum level. (d) Simulated power conversion efficiency of all type-II heterostructures studied in this work. The background with different color regions refers to different levels of PCE (color bar on the right). The heterostructures with PCE higher than 20% are labeled in this figure as "A/B", where A represents the donor and B is the acceptor. Reprinted with permission from Linghu *et al*.[13] Copyright 2018 American Chemical Society.

## 5. OTHER WORKS RELATED TO THE 2DMATPEDIA PLATFORM

Finally, we briefly summarize some recent works, utilizing the datasets in 2DMatPedia. Using the similar "bottom-up" approach in 2DmatPedia, Sorkun *et al*. used machine learning to develop a "virtual" 2D materials database with 316,505 compounds.[46] The accuracy of the predicted band gaps with ML models is tested and validated by 2DMatPedia datasets.[46] From Materials Cloud, Karmodak *et al*. only screened out 15 promising HER catalysts from 258 2D materials with $E_{exf} <$ 35 meV/Å$^2$.[47] To make a deeper data mining, Wu *et al*. used a deep learning method (at near-DFT accuracy) to screen out 38 high performance HER catalysts from 6,531 2D materials.[37]

The electronic structures are the main part of 2D materials databases. By combining 2DMatPedia with C2DB, Liang *et al.* developed an rational machine-learning model about the band gap and exciton binding energy of 2D materials, which can reveal the physics pictures behind the predicted quantity by machine learning.[48] Liu *et al.*, via screening the 2DMatPedia database, identified 15 monolayer atomic crystals hosting the unique topological flat-band feature near the Fermi level.[49] Schleder *et al.* identified 15 new topological 2D materials by screening 2D materials databases with ML algorithms.[50] Kabiraj *et al.* developed a ML model to predict the ferromagnetism of 2D materials with the test set of new and complex compositions in 2DMatPedia.[51] Lu *et al.* developed active learning by integrating ML with high-throughput calculations, which first learns ferromagnetic materials from C2DB, and then predicts 542 ferromagnetic and 917 antiferromagnetic materials from Materials Cloud and 2DMatPedia databases.[52] Leveraging the van der Waals interaction between 2D heterostructures for novel solid lubricant and super-lubricant applications, Fronzi *et al.* used machine learning approach to create a dataset of the interlayer energy and the elastic constant of 18 million van der Waals heterostructures based on 6,000+ monolayer 2D materials in 2DMatPedia.[53] Data analytic and applications based on 2DMatPedia can be expected to increase rapidly.

## 6. CONCLUSIONS AND OUTLOOK

In summary, we aim to develop capability in 2D materials informatics, revolutionize new 2D materials discovery, and shorten the time from the discovery to applications, enabled by the development of the world's largest comprehensive 2D materials database (2DMatPedia), high-throughput screening and ML models, web toolkits, and device applications. Our next progress is to i) further increase the collection of candidates in 2DMatPedia through bottom-up approach, alloying, forming heterostructures, and collection of 2D materials data from public literatures, ii) adapt advanced AI algorithms in analysis of a variety of materials properties, to find descriptors, for accelerating design of new 2D materials, and iii) utilize the platform to screen desired 2D materials for specific applications in close collaboration with experimentalists.

Despite the significant progress and successful deployment of materials genome, data-driven materials discovery, high-throughput calculations, and machine learning as a major game change in the area of 2D materials science, future challenges remain:

(1) Different research groups may create their own databases containing the calculated properties of existing and hypothetical 2D materials, such as JARVIS, C2DB, Materials Cloud, and 2DMatPedia. The collaboration and integration of them will be of great helpfulness to the researchers and users in the field of 2D materials science. Thanks to OPTIMADE, a user-friendly materials-design platform, it, recently, has integrated several materials databases (including 2DMatPedia) (https://www.optimade.org/providers-dashboard/).

(2) New complex 2D materials may not be discovered by either the top-down or bottom-up approach, such as recently synthesized $MnBi_2Te_4$. AI-enabled literature mining to collect 2D materials and their properties from public domain is an important step of building a comprehensive 2D materials database.[54,55]

(3) High-throughput calculations have the so-called speed-accuracy trade-off. Most large computed materials databases are constructed using the conventional DFT functionals for computational efficiency, which have well-known issues in the strong correlated materials, van der Waals heterostructures, and band-gap materials. This need to be addressed by methodological advances in the future.

(4) Limited by the DFT core, the high-throughput calculations still cannot handle a very large number of samples, such as the catalysis with complex reaction paths and active reaction sites. One solution is to apply active learning, a machine-learning-assisted DFT calculation.[23] The other way is to develop models to access beyond-DFT scales at near-DFT accuracy, such as materials/physics-based deep learning algorithms.[26]

(5) AI tools are being eyed with suspicion by some scientists because of the "black box" models which to be as opaque to our understanding as the data patterns themselves. The interpretable AI models for materials science are strongly desired in the future. The materials science, physics and chemistry community may learn from the ML community to transfer and implement the advanced algorithms into their domain with discipline interpretability. A recent work led by Jeffery C. Grossman in MIT is a good example along this direction. They implemented graph convolutional neural networks, a widely used algorithm for computer vision, for successfully explainable prediction of materials properties.[56]

(6) Finally, the computational community has much to collaborate with the experimental community closely. Any new discovered or designed candidate 2D materials for high-performance applications must be verified and validated by the experiment.[23] The experimental feedback inversely can be used to optimize the theoretical models to provide more accurate calculations and predictions.

We believe that future efforts in the computational 2D materials community are surely not limited to the 6 points listed above. The family of 2D materials is continuously growing both in terms of variety and quantity, and almost every new 2D material has unique properties for one or more applications. The rapid development in this field will finally accelerate the progress of novel 2D materials from the discovery to deployment in new technologies.

AUTHOR INFORMATION

**Corresponding Authors**
**Lei Shen** − *Department of Mechanical Engineering, National University of Singapore, Singapore 117575, Singapore*; orcid.org/0000-0001-6198-5753;
Email: shenlei@nus.edu.sg
**Yuan Ping Feng** − *Department of Physics, National University of Singapore, Singapore 117551, Singapore*; orcid.org/0000-0003-2190-2284;
Email: phyfyp@nus.edu.sg
**Ming Yang** − *Department of Applied Physics, The Hong Kong Polytechnic University, Hung Hom, Kowloon, Hong Kong SAR, China*; orcid.org/0000-0002-0876-1221;
Email: mingyang@polyu.edu.hk

**Authors**
**Jun Zhou** − *Institute of Materials Research and Engineering, Agency for Science, Technology and Research (A*STAR), Innovis 138634, Singapore*; orcid.org/0000-0002-5505-7616;
**Tong Yang**[†] − *Department of Physics, National University of Singapore, Singapore 117551, Singapore*; orcid.org/0000-0003-1474-1515

**Present Addresses**
[†]*Department of Applied Physics, The Hong Kong Polytechnic University, Hung Hom, Kowloon, Hong Kong SAR, China.*

**Notes**
The authors declare no competing financial interest.

Biographies

**Lei Shen** received his B.S. and M.S. degrees from Xiamen University. He is currently a senior lecturer at Department of Mechanical Engineering, National University of Singapore, Singapore. His research mainly focuses on the DFT calculations and machine learning of 2D functional materials and topological materials.
**Jun Zhou** received Ph.D. degree from National University of Singapore. He is currently a scientist at Institute of Materials Research and Engineering, A*STAR. His research mainly focuses on data-driven material discovery on 2D magnetism and high-entropy ceramics.

**Tong Yang** received his B.S. and Ph.D. degrees from Sichuan University and the National University of Singapore, respectively. He is currently a postdoctoral fellow at the Hong Kong Polytechnic University. His research mainly focuses on explorations of low-dimensional functional materials.

**Ming Yang** received Ph.D. degrees from Department of Physics, National University of Singapore. He is currently an assistant professor at the Department of Applied Physics, The Hong Kong Polytechnic University. His research mainly focuses on high-throughput first-principles calculations and machine learning accelerated material development.

**Yuan Ping Feng** received his Ph.D. degrees in Physics from Illinois Institute of Technology in 1987. He is currently a Professor in Department of Physics, National University of Singapore. His research interest is in computational condensed matter & materials physics, focusing mainly on the understanding of fundamental properties of materials for advanced technologies, and prediction of new materials based on ab initio electronic structure calculations and materials genomic approach. Over the years, his group has studied various materials including structures, properties and applications of two-dimensional materials, dilute magnetic semiconductors, graphene spintronics, topological insulators, high-$k$ materials, semiconductor and metal surfaces and interfaces, materials for magnetic data storage, etc.

ACKNOWLEDGMENT

This research/project is supported by the Ministry of Education, Singapore, under its MOE AcRF Tier 2 (Award MOE2019-T2-2-030), RSB Grant (No. C-144-000-207-532), and MOE Tier 1 (Grants No. R-144-000-441-114, No. R-144-000-413-114, No. R-265-000-651-114, and No. R265-000-691-114).

TOC image

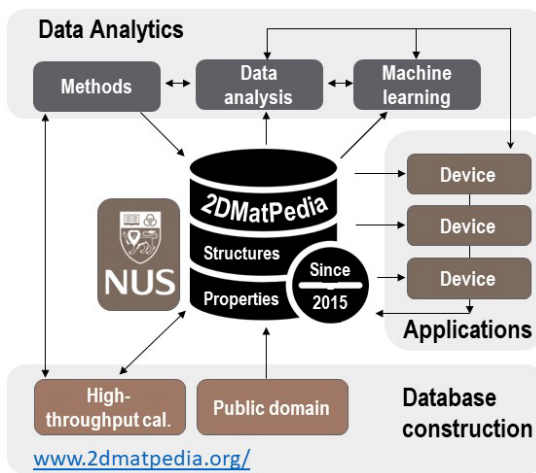